\documentclass[%
 aip,
 amsmath,amssymb,
 reprint,%
]{revtex4-1}

\usepackage{graphicx}
\usepackage{dcolumn}
\usepackage{bm}

\usepackage[utf8]{inputenc}
\usepackage[T1]{fontenc}
\usepackage{mathptmx}

\begin{document}

\preprint{AIP/123-QED}

\title{The magnetic, electronic and light-induced topological properties in two-dimensional hexagonal FeX$_2$ (X=Cl, Br, I) monolayers}

\author{Xiangru Kong}
\affiliation{International Center for Quantum Materials and School of Physics,
Peking University, Beijing 100871, China}
\affiliation{Collaborative Innovation Center of Quantum Matter, Beijing 100871, China.}
\affiliation{Center for Nanophase Materials Sciences, Oak Ridge National Laboratory,
Oak Ridge, Tennessee 37831, United States}
\author{Linyang Li}
\email{linyang.li@hebut.edu.cn}
\affiliation{School of Science, Hebei University of Technology, Tianjin 300401, China}
\affiliation{Department of Physics, University of Antwerp, Groenenborgerlaan 171,
B-2020 Antwerp, Belgium}

\author{Liangbo Liang}
\affiliation{Center for Nanophase Materials Sciences, Oak Ridge National Laboratory,
Oak Ridge, Tennessee 37831, United States}
\author{Fran\c{c}ois M. Peeters}
\affiliation{Department of Physics, University of Antwerp, Groenenborgerlaan 171,
B-2020 Antwerp, Belgium}
\author{Xiong-Jun Liu}
\affiliation{International Center for Quantum Materials and School of Physics,
Peking University, Beijing 100871, China}
\affiliation{Collaborative Innovation Center of Quantum Matter, Beijing 100871, China.}
\affiliation{Beijing Academy of Quantum Information Science, Beijing 100193, China}
\affiliation{CAS Center for Excellence in Topological Quantum Computation, University of Chinese Academy of Sciences, Beijing 100190, China}

\date{\today}

\begin{abstract}
 Using Floquet-Bloch theory, we propose to realize chiral topological phases in 2D hexagonal FeX$_2$ (X=Cl, Br, I) monolayers under irradiation of circularly polarized light.
Such 2D FeX$_2$ monolayers are predicted to be dynamical stable,
and exhibit both ferromagnetic and semiconducting properties. To capture
the full topological physics of the magnetic semiconductor under periodic driving,
we adopt \textit{ab initio} Wannier-based tight-binding methods for the
Floquet-Bloch bands, with the light-induced band gap closings and openings being obtained as the light field strength increases. The
calculations of slab with open boundaries show the existence of chiral
edge states. Interestingly, the topological transitions with branches of chiral edge
states changing from zero to one and from one to two by tuning
the light amplitude are obtained, showing that the topological Floquet
phase of high Chern number can be induced in the present Floquet-Bloch systems.
\end{abstract}

\maketitle

The discovery of the integer quantum Hall (QH) effect~\cite{PRL1980QH}, which is characterized by topological Chern number~\cite{PRL1982tknn}, brought about the new fundamental notion of topological quantum phases. The topological phases are now a mainstream of research in
condensed matter physics, with the research being greatly revived after the discovery of topological insulators~\cite{PRL2005QSH,PRL2006QSH,RMP2010TI,RMP2011TI}. 
Among the topological phases, the magnetic topological materials are of peculiar interests. In particular, the quantum anomalous Hall (QAH) effect, the version of QH effect without Landau levels and first proposed by Haldane~\cite{PRL1988qah}. was first realized in experiment by in a thin film magnetic topological insulator~\cite{chang2013qah}. The magnetic topological materials with robust topological
edge states have the potential application for future nanodevices~\cite{RMP2016TI,kou2017two,PhysRevApplied.9.054023}. Moreover, the chiral topological superconductivity with chiral Majorana modes was reported based on QAH phases in proximity of an $s$-wave superconductor~\cite{he2017chiral}. The Majorana zero modes in topological superconductor can obey non-Abelian statistics and have important potential applications to topological quantum computation~\cite{RMP2008TQC,alicea2012new,PRX2014XJ,PRL2017XJ}, and thus have also attracted considerable interests.

Nevertheless, the magnetic materials 
are rare in comparison with nonmagnetic materials~\cite{RMP2014fm,dietl2010ten}.
Magnetic materials are the major platform for spintronics~\cite{gutfleisch2011magnetic},
and the controllability of the spin degree of electrons depends on the material dimension~\cite{wolf2001spintronics,kong2014spin}. On the other hand, due to the discovery
of graphene~\cite{RMP2009gra,novoselov20162d,miro2014atlas}, two-dimensional
(2D) atomic like materials have attracted great interests. Many 2D
materials have been studied such as elemental group IV or V monolayers~\cite{miro2014atlas,mannix2017synthesis,PRB2017Bi}
and transition metal dichalcogenide (TMD) monolayers~\cite{wang2012electronics,chhowalla2013chemistry,li2014structures}.
To realize spintronics in 2D materials, a typical way is to dope
magnetic elements~\cite{RMP2014fm}, which however has the disadvantage in leading to disorders. In comparison, due to the partially filled $d$
sub-shell in transition metals~\cite{kong2018quantum}, the intrinsic magnetic semiconductors may be realized in 2D TMD or other similar transition
metal compounds such as transition metal halides and AB$_2$ type monolayers\cite{li2020high}.

Another important issue is how to manipulate the topological quantum phase. Despite a large amount of 2D topological materials that have been predicted~\cite{RMP2016TI,kou2017two}, controlling topological phase
transitions usually requires changing the structural properties of
materials~\cite{bernevig2006quantum,konig2007quantum,zhao2015strain}.
Recently, the non-equilibrium schemes for topological phases, including the quantum quenches~\cite{zhang2018dynamical} and periodic driving by optical pumping with time-dependent perturbations~\cite{claassen2016all,wang2013observation,mahmood2016selective,PhysRevB.79.081406}, enable the non-equilibrium manipulations of topological quantum phases and attracted fast growing attention.
Earlier proposal for Floquet-Bloch states were studied
in graphene~\cite{PhysRevB.79.081406,PhysRevB.89.121401,PhysRevLett.113.266801} and  HgTe/CdTe quantum wells~\cite{lindner2011floquet}. In experiments,
Floquet-Bloch states were observed in graphene~\cite{mciver2020light} and on the surface of topological insulators~\cite{wang2013observation,mahmood2016selective}.
Due to the existence of electronic valley degree of freedom in TMD~\cite{schaibley2016valleytronics},
It is interesting to investigate Floquet physics in TMD~\cite{Sie2014ws2,claassen2016all,de2016monitoring},
and study the topological phases in magnetic semiconducting transition
metal compounds by Floquet-Bloch theory~\cite{sentef2015theory,PRL2018BP,hubener2017creating,PRL2013FB,PRL2018FeSe,PRB2011fbg}.

In this work, after showing that hexagonal (2H) FeX$_2$ (X=Cl, Br, I) monolayers
are ferromagnetic semiconductors that are dynamically stable, we propose to realize and control the topological phases in FeX$_2$ by coupling it to circularly polarized light. For the present Floquet-Bloch
states, instead of using a simple band description which may fail,
we use \textit{ab initio} Wannier-based tight-binding methods for
calculating the Floquet-Bloch band structures and the light-induced
chiral edge states. Different Floquet topological states with high Chern number are obtained by tuning the chirality and strength of circularly polarized light, showing the intriguing controllability of the present topological matter.

First-principles calculations were implemented in the Vienna \textit{ab
initio} simulation package (VASP) with projector augmented wave (PAW)
method~\cite{PRB-1996-iterative-PW,PRB-1999-PAW} in the framework
of Density Functional Theory (DFT)~\cite{PhysRev-1964-DFT}. The electron
exchange-correlation functional was adopted as the generalized gradient
approximation (GGA) in the form proposed by Perdew, Burke and Ernzerhof
(PBE)~\cite{PRL-1996-PBE}. The structural relaxation considering
both the atomic positions and lattice vectors was performed by using
the conjugate gradient (CG) scheme until the maximum force on each
atom was less than 0.01 eV/$\mathring{A}$, and the total energy is converged
to $10^{-5}$ eV. The energy cutoff of the plane waves was chosen
as 520 eV. The Brillouin zone (BZ) integration was sampled by using
a $\Gamma$-centered 21 $\times$21 $\times$ 1 Monkhorst-Pack grid.
To study the dynamical stability of 2H FeX$_2$ (X=Cl, Br, I) monolayers, phonon frequencies were calculated by the finite displacement method
with the \textit{Phonopy} code~\cite{phonopy}. The Floquet-Bloch
band structures, slab and surface state calculations were illustrated
with \textit{ab initio }Wannier-based tight-binding methods~\cite{Mostofi20142309,WU2017}.
The iterative Green's function method was used for surface state calculations~\cite{sancho1985highly}.
The Wilson loop or Wannier Charge Centers (WCCs) method was used for the calculations of Chern number~\cite{PhysRevB.83.235401,PhysRevB.84.075119,PhysRevB.95.075146}.

\begin{figure}
\begin{centering}
\includegraphics[scale=0.4]{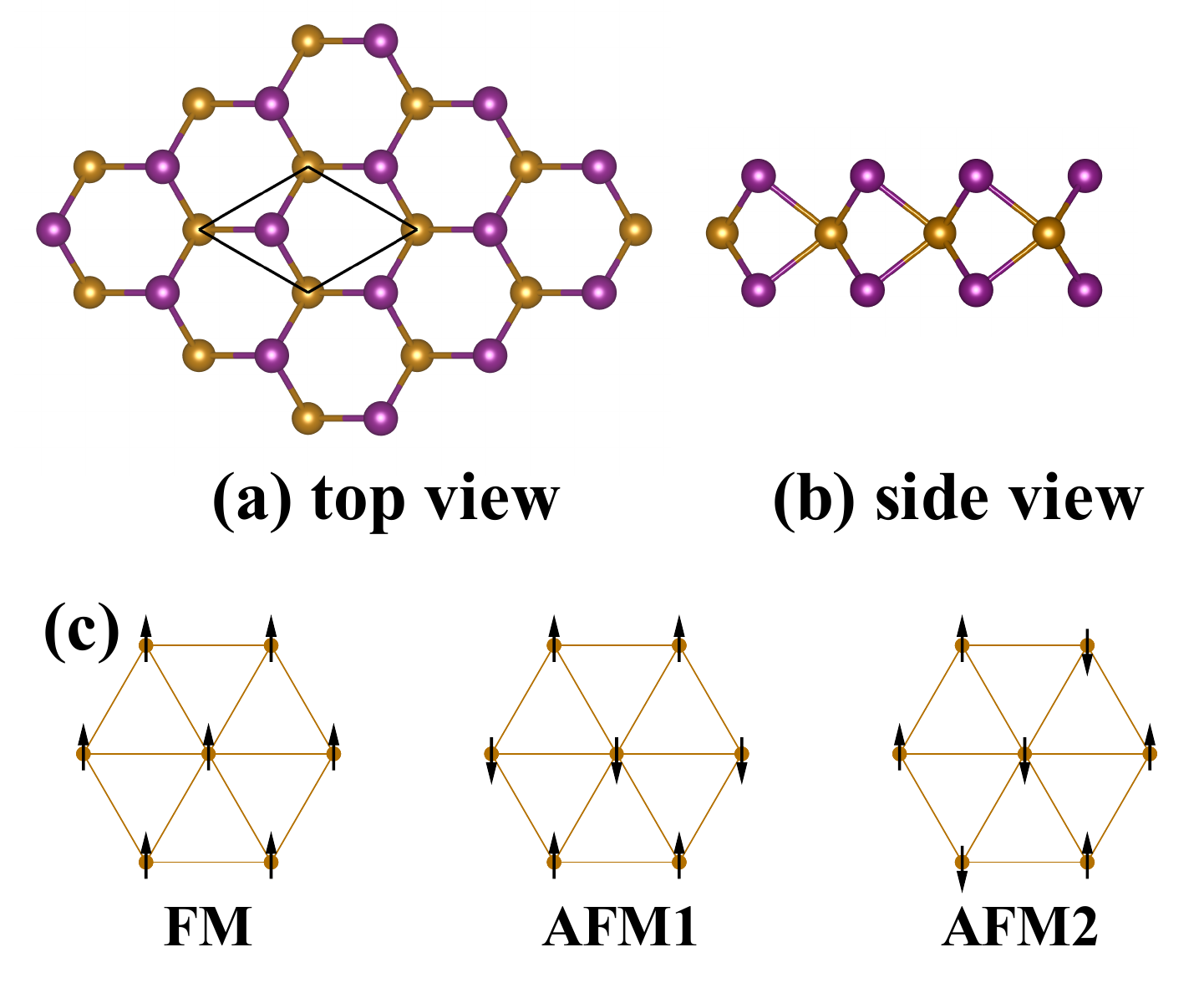}
\par\end{centering}
\caption{\label{fig:structure}The lattice structure of FeX$_2$: (a) top view
and (b) side view. (c) Three magnetic configurations: FM, AFM1, AFM2.
The golden symbols indicate the Fe atom, and the purple symbols refer
to the halogen atom (X=Cl, Br, I). FM = ferromagnetic, AFM = antiferromagnetic.}

\end{figure}

The hexagonal (2H) FeX$_2$ (X=Cl, Br, I) monolayers possess the point
group $D_{3h}$ with broken inversion symmetry, which is the same
as in 2H TMD monolayers~\cite{wang2012electronics,chhowalla2013chemistry}.
Honeycomb structure is shown from the top view (Fig. \ref{fig:structure}(a)),
and sandwich-like stacking of atom layers (X-Fe-X) is observed from
the side view (Fig. \ref{fig:structure}(b)). Recently, the triangular
(1T) FeX$_2$ monolayers were predicted to exhibit ferromagnetic (FM)
ground state with metallic electronic properties~\cite{torun2015stable,ashton2017two}.
Here, we investigate the magnetic ground states of 2H FeX$_2$ monolayers
by constructing a 2$\times$2$\times$1 supercell. Three magnetic
configurations are considered as shown in Fig. \ref{fig:structure}(c):
FM, AFM1, AFM2. The energy calculations indicate that the magnetic
ground state of 2H FeX$_2$ monolayers is FM. Taking FeI$_2$ for example,
the energy of FM state is about 125 meV per Fe atom lower than that
of the AFM1 state, and about 159 meV per Fe atom lower than that of
the AFM2 state. The Fe ($3d^{6}4s^{2}$) atom donates one electron
to each halogen atom (X=Cl, Br, I) such that the new electron configuration
of Fe atom becomes $3d^{4}4s^{2}$, and the four electrons in the
$d$ orbital will form $\uparrow\uparrow\uparrow\uparrow$ spin configuration.
DFT calculations confirm that the total magnetization of the 2H FeX$_2$
monolayers in the unit cell is 4 $\mu_{B}$. By DFT magnetic
calculations, the easy magnetic axis of 2H FeX$_2$ is in-plane, and
the Magnetocrystalline Anisotropy Energy (MAE) of the out-of-plane
spin orientation which can be aligned by magnetic field is about 1.1
meV per spin for every Fe atom; the total energy of different directions of spin orientation is shown in Fig.~S10 in the Supporting Information
(SI). To demonstrate the dynamical stability
of FM 2H FeX$_2$ monolayers, the phonon spectrum without imaginary
frequency modes are shown in Fig. S1.

\begin{figure}
\begin{centering}
\includegraphics[scale=0.08]{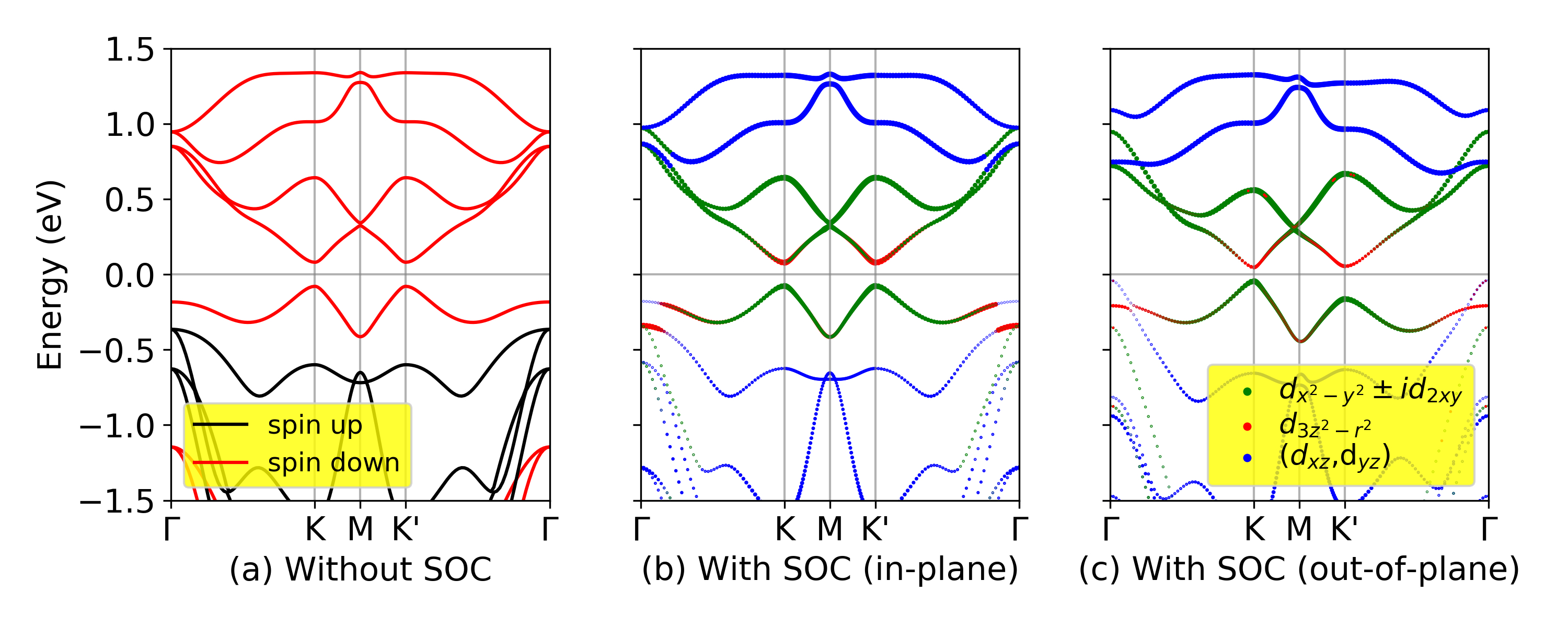}
\par\end{centering}
\caption{\label{fig:bandstructures}The band structures of FeI$_2$: (a) magnetic state without SOC; (b) in-plane magnetic state
(with SOC); (c) out-of-plane magnetic state (with SOC).}

\end{figure}

\begin{figure*}[htb!]
\begin{centering}
\includegraphics[scale=0.4]{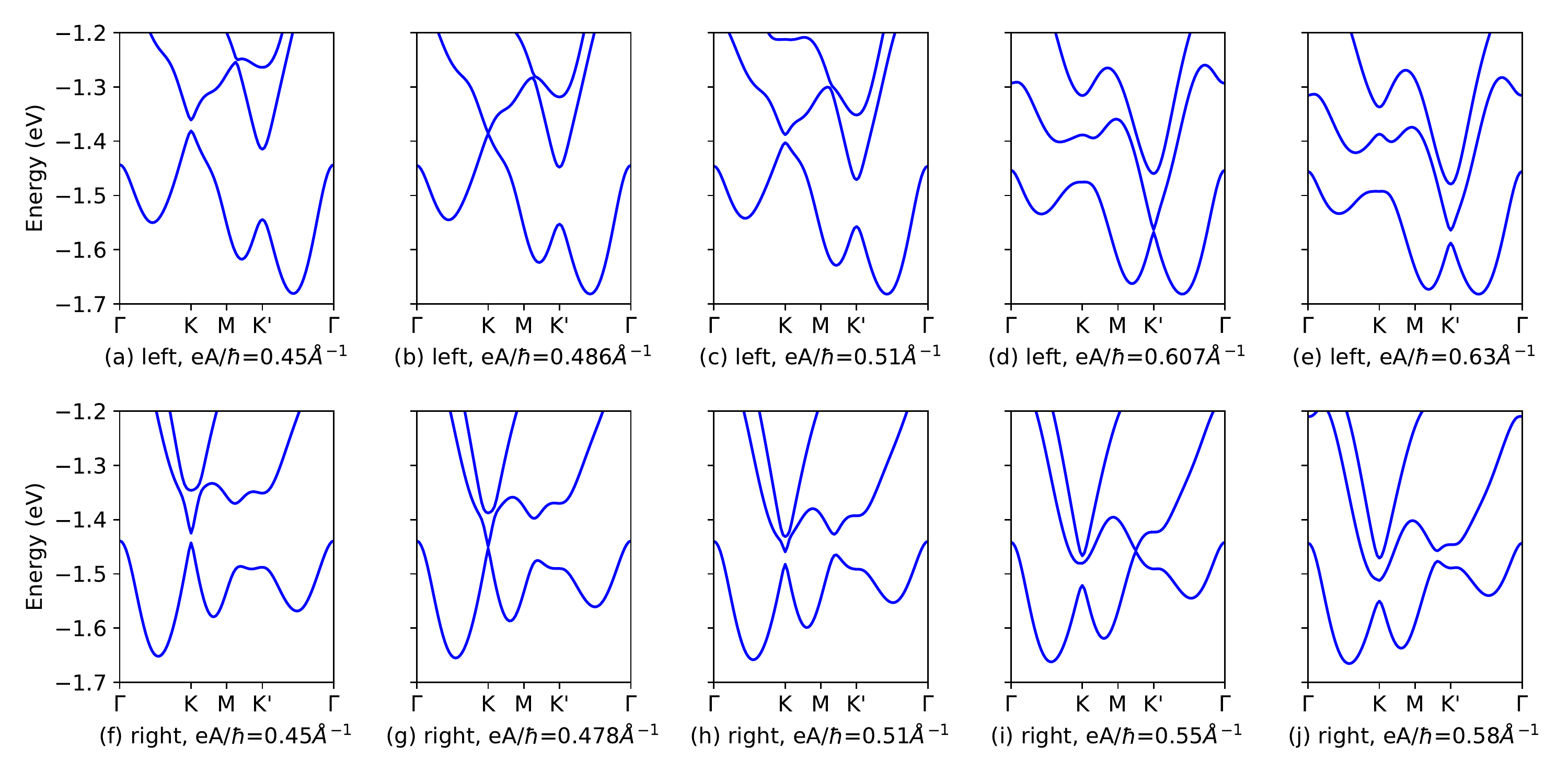}
\par\end{centering}
\caption{\label{fig:FBband}The quasi-energy spectrum of FeI$_2$ with
out-of-plane magnetization under the irradiation of (a)-(e) left and
(f)-(j) right circularly polarized light.}

\end{figure*}

Here, we consider the electronic band structures of FeX$_2$ monolayers
for magnetic state without spin-orbit coupling (SOC),
and magnetic state with SOC (in-plane and out of plane
magnetization). The magnetic state without SOC of FeI$_2$ exhibits
spin-polarized bands as shown in Fig. \ref{fig:bandstructures}(a),
and similar results can be found for FeCl$_2$ in Fig. S2 and FeBr$_2$ in Fig. S3. The
bands near the Fermi level consists mostly of transition-metal \textit{d}-orbitals,
which could be split into three groups, $d_{3z^{2}-r^{2}}$, $d_{x^{2}-y^{2}}\pm id_{2xy}$
and $d_{xz}\pm id_{yz}$ based on the symmetry analysis as shown
in Figs. 2(b) and (c) (also see Figs. S2 and S3)~\cite{claassen2016all}.
Considering the atom radius and electronegativity of halogen atoms (Cl, Br, I), the
lattice constants ($a$), the height ($h$) and the Fe-X bond length
($d$) of the FeX$_2$ monolayer increase (see Table SI).
This results in weaker repulsive interaction between the $d$ orbitals
of the Fe atoms, and consequently the band gaps between the different
group of $d$ orbitals become smaller. Therefore, the band gaps between
$d_{3z^{2}-r^{2}}$ and $d_{x^{2}-y^{2}}\pm id_{2xy}$ in the magnetic calculations without SOC are 395.4, 268.1 and 159.8 meV for FeCl$_2$,
FeBr$_2$ and FeI$_2$, respectively. As the space group of 2H FeX$_2$
monolayer is $P\bar{6}m2$, the generators of this space group are
$C_{3z}$ (three-fold rotational symmetry along the $z$ direction),
$m_{z}$ (out-of-plane mirror symmetry) and $m_{110}$ (in-plane mirror
symmetry). The in-plane magnetization with SOC preserves the symmetry,
so the energy of in-plane magnetic calculations with SOC along
K-M and M-K' keeps almost degenerate which results in the same band
gaps at K and K', as shown in Figs. \ref{fig:bandstructures}(b), S2(b) and
S3(b). However, the out-of-plane magnetization violates
the $m_{z}$ symmetry, SOC effect will break the degeneracy along M-K' and K-M\cite{zhaos2020intrinsic}
and thus the band gaps at K and K' have different values as shown in Figs.
\ref{fig:bandstructures}(c), S2(c), S3(c) and Table SI.
The average band gap of out-of-plane magnetization at K and K' is
about the same as that of in-plane magnetization. On the other hand,
different SOC strength of halogen atoms brings about the change of
band gaps ($\delta$) as 1.0, 1.7 and 4.3 meV for FeCl$_2$, FeBr$_2$
and FeBr$_2$, respectively (see Table SI).

Here, the Wannier orbitals are projected on the $d$ orbitals in Fe
atom and $p$ orbitals in I atoms in the \textit{ab initio }Wannier-based
tight-binding methods. An external time-dependent circularly polarized
light $A(t)=A(\eta sin(\omega t),cos(\omega t),0)$ irradiates the
FeI$_2$ monolayer along the $z$ direction, where $\omega$ is the
frequency, $\eta=\pm1$ indicates the chirality of the circularly
polarized light, and $A$ is the light amplitude. By minimal coupling~\cite{PRL2013FB},
the time-dependent tight-binding Hamiltonian becomes
\[
H(k,t)=\sum_{m,n}\sum_{j}t_{j}^{mn}(t)e^{ik\cdot R_{j}}c_{m}^{\dagger}(k,t)c_{n}(k,t),
\]
where $t_{j}^{mn}(t)=t_{j}^{mn}e^{i\frac{e}{\hbar}A(t)\cdot d_{j}^{mn}}$
with $t_{j}^{mn}$ is the hopping term, $R_{j}$ is the lattice vector
in the $j$-cell, and $d_{j}^{mn}$ is the position vector between two
Wannier orbitals $m$ and $n$. According to Floquet-Bloch theory~\cite{PRL2018BP,PRL2013FB,PRL2018FeSe,PhysRevB.82.235114,PRB2011fbg},
an effective static Hamiltonian can be
\[
H_{F}(k)=\sum_{m,n}\sum_{\alpha,\beta}[h_{\alpha-\beta}^{mn}(k)+\alpha\hbar\omega\delta_{mn}\delta_{\alpha\beta}]c_{\alpha m}^{\dagger}(k)c_{\beta n}(k),
\]
\[
h_{\alpha-\beta}^{mn}=\sum_{j}t_{j}^{mn}\cdot\frac{1}{T}\int_{0}^{T}dte^{i[\frac{e}{\hbar}A(t)\cdot d_{j}^{mn}+(\alpha-\beta)\omega t]}\cdot e^{ik\cdot R_{j}},
\]
where  $\text{\ensuremath{\hbar\omega=8eV}}$  is chosen to be larger than the band width of FeI$_2$, so that the Floquet bands do not cross each other and this high frequency regime could be could be reached by F$_2$ excimer laser\cite{white1984submicron}; and in this high frequency regime $\alpha-\beta=\pm1$ can be used to get  converged results in our work without considering more Floquet bands~\cite{PRL2013FB}.

As shown in Fig. \ref{fig:FBband}, the Floquet-Bloch band structures
of out-of-plane magnetization behave differently under different chirality
of circularly polarized light, and this is similar with valley-dependent optical selection rules in  inversion symmetry breaking system~\cite{yao2008valley}. As the amplitude $A$ increases, the
circularly polarized light will induce gap closing and opening at
the K and K' points. Under irradiation of left circularly polarized
light, as $eA/\hbar$ increases to about 0.486 $\mathring{A}^{-1}$, the first
gap closing at K point appears as shown in Fig. \ref{fig:FBband}(b);
as $eA/\hbar$ increases to about 0.607 $\mathring{A}^{-1}$, the gap at K
point enlarges, but the gap at the K' point closes as shown in Fig.
\ref{fig:FBband}(d); then as $eA/\hbar$ increases further, both
of the gaps at the K and K' enlarges as shown in Fig. \ref{fig:FBband}(e).
For right circularly polarized light, the first gap closing appears
at K point when $eA/\hbar=0.478\mathring{A}^{-1}$ (Fig. \ref{fig:FBband}(g)),
and the second gap closing appears at K' point when $eA/\hbar=0.55\mathring{A}^{-1}$
(Fig. \ref{fig:FBband}(i)) which is less than that of left circularly
polarized light. Different from the out-of-plane magnetization where
the degeneracy of band energies is broken at K and K' points, the
response of the in-plane magnetization to the time-dependent left or right circularly polarized
light is similar at K and K' points
as shown in Fig. S4. Under irradiation of left (right) circularly
polarized light, the first gap closing appears at K' (K) point when
$eA/\hbar=0.55\mathring{A}^{-1}$ as shown in Fig. S4 (b) (Fig. S4 (g)), and
the second gap closing appears at K (K') when $eA/\hbar=0.57\mathring{A}^{-1}$
as shown in Fig. S4 (d) (Fig. S4 (i)).

\begin{figure}
\begin{centering}
\includegraphics[scale=0.17]{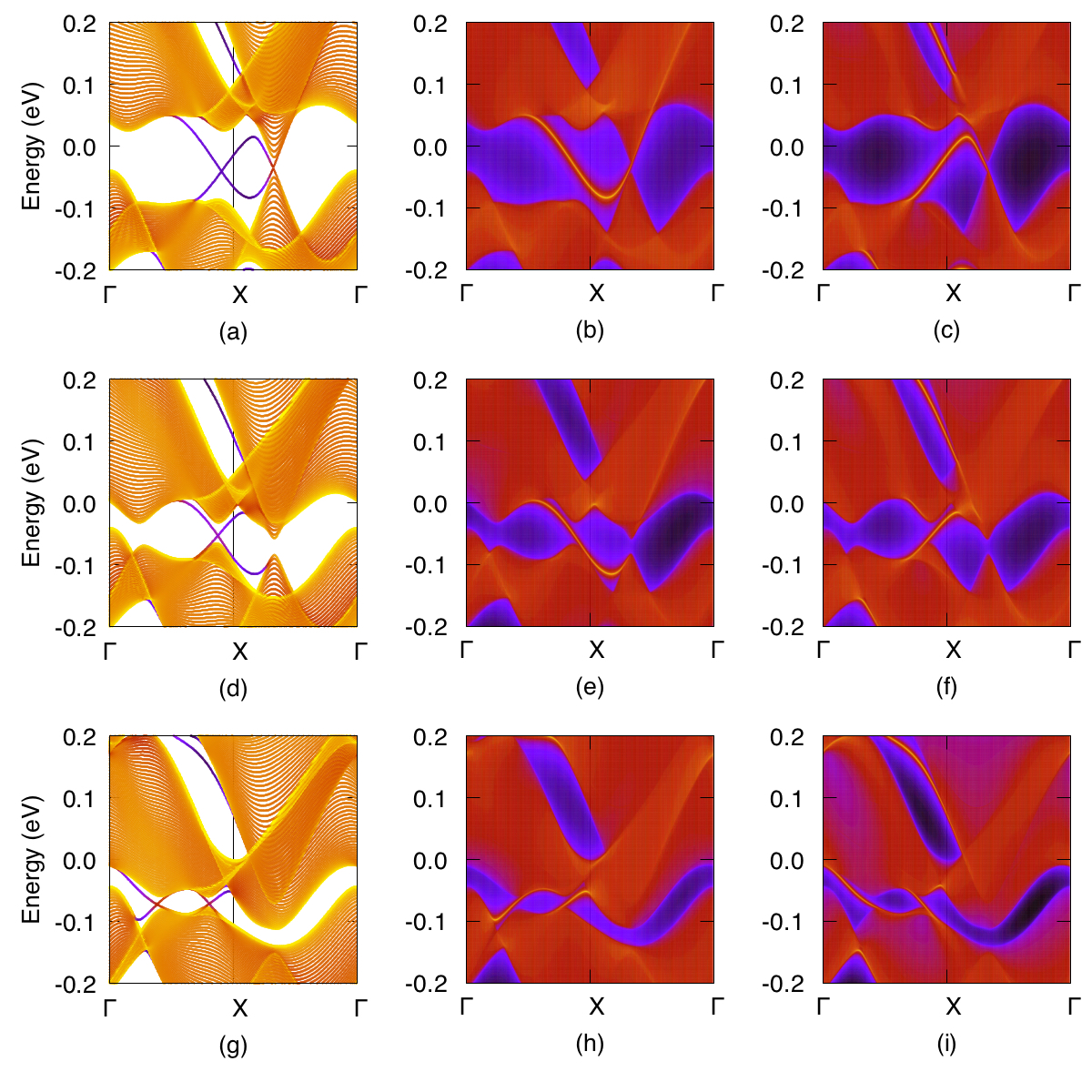}
\par\end{centering}
\caption{\label{fig:edge}The quasi-energy spectrum of FeI$_2$ with out-of-plane
magnetization under the irradiation of right circularly polarized
light for slab and edge states calculations. (a)-(c) $eA/\hbar=0.45\mathring{A}^{-1}$,
(d)-(f) $eA/\hbar=0.51\mathring{A}^{-1}$ and (g)-(i) $eA/\hbar=0.58\mathring{A}^{-1}$.
The left column shows the slab calculations, the middle column shows
the left edge states calculations and the right column shows the right
edge states calculations.}  

\end{figure}

Gap closing and opening always indicate topological phase transition
which is complementary to the Laudau symmetry broken theory~\cite{RMP2010TI,RMP2011TI}.
Bulk-edge correspondence is an important reflection of topological
phases, e.g., nontrivial edge states will appear on the edges of nanoribbons
of 2D topological materials. According to the above discussions on
the Floquet-Bloch band structures, there are two gap closings as the
light amplitude increases. Under the irradiation of right circularly
polarized light and for out-of-plane magnetization, we choose $eA/\hbar=0.45\mathring{A}^{-1}$,
$0.51\mathring{A}^{-1}$, and $0.58\mathring{A}^{-1}$ for the slab and edge states
calculations as shown in Fig. \ref{fig:edge}. Before the first gap
closing, there are trivial edge states appearing in the gap between
the valence and conduction bands as shown in Fig. \ref{fig:edge}(a);
one of the edge states connects only the conduction (valence) bands
on the left (right) edge as shown in Fig. \ref{fig:edge}(b) or (c),
which indicates the trivial properties of the edge states; here we track the evolution of the sum of WCCs as shown in Fig. S8(a), no winding in one periodic path indicates Chern number $C=0$. After the first gap closing at K point when $eA/\hbar=0.51\mathring{A}^{-1}$, two counterpropagating
nontrivial edge states could be observed in Fig. \ref{fig:edge}(d):
there is one left (right) propagating nontrivial edge state connecting
valence and conduction bands on the left (right) edge as shown in
Fig. \ref{fig:edge}(e) (Fig. \ref{fig:edge}(f)); the evolution of the sum of WCCs as shown in Fig. S8(b) indicates Chern number $C=-1$ which the downward winding gives the minus value. In the off-resonant driving paradigm with high frequency, the driving filed will cause optical Stark shift of different magnitude on K and K' in the Floquet bands. This results in the band inversions with the increasing amplitude of light. Put another way, the driving filed will induce a “mass term” to result in the band inversions in the Floquet bands\cite{rudner2019floquet}. Thus it will cause the trivial edge state that only connecting valence or conduction band to become one chiral edge state that connect valence and conduction bands.

Surprisingly, after the second gap closing at K' point, one additional nontrivial edge
state appears on each edge as shown in Fig. \ref{fig:edge}(g), which
indicates the transition between two different nontrivial topological
phases. Furthermore, the propagating direction of two nontrivial
edge states on each edge is reversed from that of topological phase
with one nontrivial edge, as shown in Figs. \ref{fig:edge}(h) and
(i); importantly, this is confirmed by the twice upward winding in the loop by tracking the evolution of the sum of WCCs as shown in Fig. S8(c), which gives Chern number $C=2$. Under the irradiation of left circularly polarized light and
for out-of-plane magnetization, trivial and nontrivial edge states
appear on the edges of the monolayer before and after the gap closing
as shown in Figs. S5(a)-(f), which is similar to that of out-of-plane
magnetization under the irradiation of right circularly
polarized light; however, because the gaps at K and K' points appear in rather
large energy ranges, the edge states annihilate in the bulk states
as shown in Figs. S5(g)-(i). As for the in-plane magnetization (Figs.
S6 and S7) under both chirality of circularly polarized light, after
the first gap closing, there are nontrivial edge states appearing;
but after the second gap closing, nontrivial edge states no longer
appear, and a phase transition from nontrivial to trivial happens
which is quite different from that of out-of-plane magnetization.
The phase diagram of all conditions are summarized in Fig. S9.

The stable 2H FeX$_2$ (X=Cl, Br, I) monolayers and their intrinsic
ferromagnetism enrich the family of magnetic materials, having the
flexible controllability inherent to 2D magnetic materials. The semiconducting
properties have potential applications in spintronics. Based on \textit{ab
initio }Wannier-based tight-binding methods, we fully capture the
electronic band structural details of 2H FeI$_2$ monolayer and investigate
the Floquet-Bloch states by Floquet-Bloch theory. Left and right circularly
polarized light for in-plane and out-of-plane magnetization were considered,
resulting in different topological behaviors. The Floquet-Bloch band structures
demonstrate the light-induced band gap closing and opening. More interestingly, under the irradiation
of right circularly polarized light and for out-of-plane magnetization,
the increasing light amplitude will not only induce a phase transition
from trivial to one nontrivial chiral edge state, but also induce
a topological phase transition with the number of chiral edge states
from one to two. The interesting phenomena in 2H FeX$_2$ monolayers
hold the promise for future applications in nanodevices.

See supplementary material for the lattice parameters, electronic band structures and Floquet-Bloch band structures of 2H FeX$_2$ (X=Cl, Br, I) monolayers in other cases and further discussions.

Data available on request from the authors~\cite{data}.

\begin{acknowledgments}
This work was supported by Ministry of Science and Technology of China
(MOST) (Grant No. 2016YFA0301604), National Natural Science Foundation
of China (NSFC) (No. 11574008, 11761161003, 11825401, and 11921005), Strategic Priority Research Program of Chinese Academy of Science
(Grant No. XDB28000000), the
Fonds voor Wetenschappelijk Onderzoek (FWO-Vl), and the FLAG-ERA Project
TRANS 2D TMD. The computational resources and services used in this
work were provided by the VSC (Flemish Supercomputer Center), funded
by the Research Foundation - Flanders (FWO) and the Flemish Government
- department EWI, and the National Supercomputing Center in Tianjin,
funded by the Collaborative Innovation Center of Quantum Matter.
This research also used resources of the Compute and Data Environment for Science (CADES) at the Oak Ridge National Laboratory, which is supported by the Office of Science of the U.S. Department of Energy under Contract No. DE-AC05-00OR22725.
X.K. and L.L. also 
acknowledge work conducted at the Center for Nanophase Materials Sciences, which is a US Department of Energy Office of Science User Facility.
\end{acknowledgments}

\bibliographystyle{apsrev4-1}
\bibliography{reference}

\end{document}